\documentclass[twocolumn,showpacs,preprintnumbers,amsmath,amssymb, superscriptaddress, prb]{revtex4-1}
\usepackage{amsmath}
\usepackage{graphicx}
\usepackage{dcolumn}
\usepackage{bm}
\usepackage[caption=false]{subfig}
\usepackage[bottom]{footmisc}
\usepackage{color}

\begin{document}
\preprint{APS}

\author {Mariano de Souza}
\email{mariano@rc.unesp.br}
\affiliation{S\~ao Paulo State University (UNESP), IGCE, Rio Claro, SP, Brazil}
\author {Lucas Squillante}
\affiliation{S\~ao Paulo State University (UNESP), IGCE, Rio Claro, SP, Brazil}
\author {Cesar S\^onego}
\affiliation{S\~ao Paulo State University (UNESP), IGCE, Rio Claro, SP, Brazil}
\author {Paulo Menegasso}
\affiliation{S\~ao Paulo State University (UNESP), IGCE, Rio Claro, SP, Brazil}
\author{Pascale Foury-Leylekian}
\affiliation{Laboratoire de Physique des Solides, CNRS UMR 8502,  Univers.\,Paris Sud, Universit\'e Paris Saclay, Orsay, France}
\author{Jean-Paul Pouget}
\affiliation{Laboratoire de Physique des Solides, CNRS UMR 8502,  Univers.\,Paris Sud, Universit\'e Paris Saclay, Orsay, France}


\title{Probing the Ionic Dielectric Constant Contribution in the Ferroelectric Phase of the Fabre-Salts}

\vspace{0.5cm}

\begin{abstract}
In strongly correlated organic materials it has been pointed out that charge-ordering could also achieve electronic ferroelectricity at the same critical temperature $T_{co}$. A prototype of such phenomenon are the quasi-one dimensional (TMTTF)$_2X$ Fabre-salts. However, the stabilization of a long-range ferroelectric ground-state below $T_{co}$ requires the break of inversion symmetry, which should be accompanied by a lattice deformation. In this work we investigate the role of the monovalent counter-anion $X$ in such mechanism. For this purpose, we measured the quasi-static dielectric constant along the $c^{*}$-axis direction, where layers formed by donors and anions alternate. Our findings show that the ionic charge contribution is three orders of magnitude lower than the intra-stack electronic response. The $c^{*}$ dielectric constant ($\epsilon'_{c^*}$)  probes directly the charge response of the monovalent anion $X$, since the anion mobility in the structure should help to stabilize the ferroelectric ground-state. Furthermore, our $\epsilon'_{c^*}$ measurements 
show that the dielectric response is thermally broaden below $T_{co}$ if the ferroelectric transition occurs in the temperature range where the anion movement begin to freeze in their methyl groups cavity. In the extreme case of the PF$_6$-H$_{12}$ salt, where $T_{co}$ occurs at the freezing point, a relaxor-type ferroelectricity is observed. Also, because of the slow kinetics of the anion sub-lattice, global hysteresis effects and reduction of the charge response upon successive cycling are observed. In this context, we propose that anions control the order-disorder or relaxation character of the ferroelectric transition of the Fabre-salts. Yet, our results show that x-ray irradiation damages change the well-defined ferroelectric response of the AsF$_6$ pristine salt into a relaxor.
\end{abstract}

\maketitle
\vspace{1.0cm}

\date{\today}

\section{Introduction}
Currently, there is a considerable interest both for fundamental and application purposes in the study of electronic systems exhibiting strong electron-electron interactions. Electronic correlations lead to subtle interplay between charge, lattice and spin degrees of freedom, whose most illustrative examples can be found among organic metals \cite{Lebed}. As a consequence, the phase diagram of such organic metals exhibits a subtle competition between spin or charge modulated ground-states and superconductivity \cite{Lebed,reviewdressel,MdS2013}. Among the various charge modulated ground-states, a ferroelectric phase has been recently discovered in several families of organic materials \cite{Review1,Review2,Ishihara}. This discovery has boosted the study of ferroelectric materials because the origin of ferroelectricity in the organics is basically electronic \cite{Brazo,BrazoII,BrazoIII}, in contrast to the ionic origin in conventional ferroelectric systems studied for decades.

\begin{figure}
\centering
\includegraphics[angle=0,width=\columnwidth]{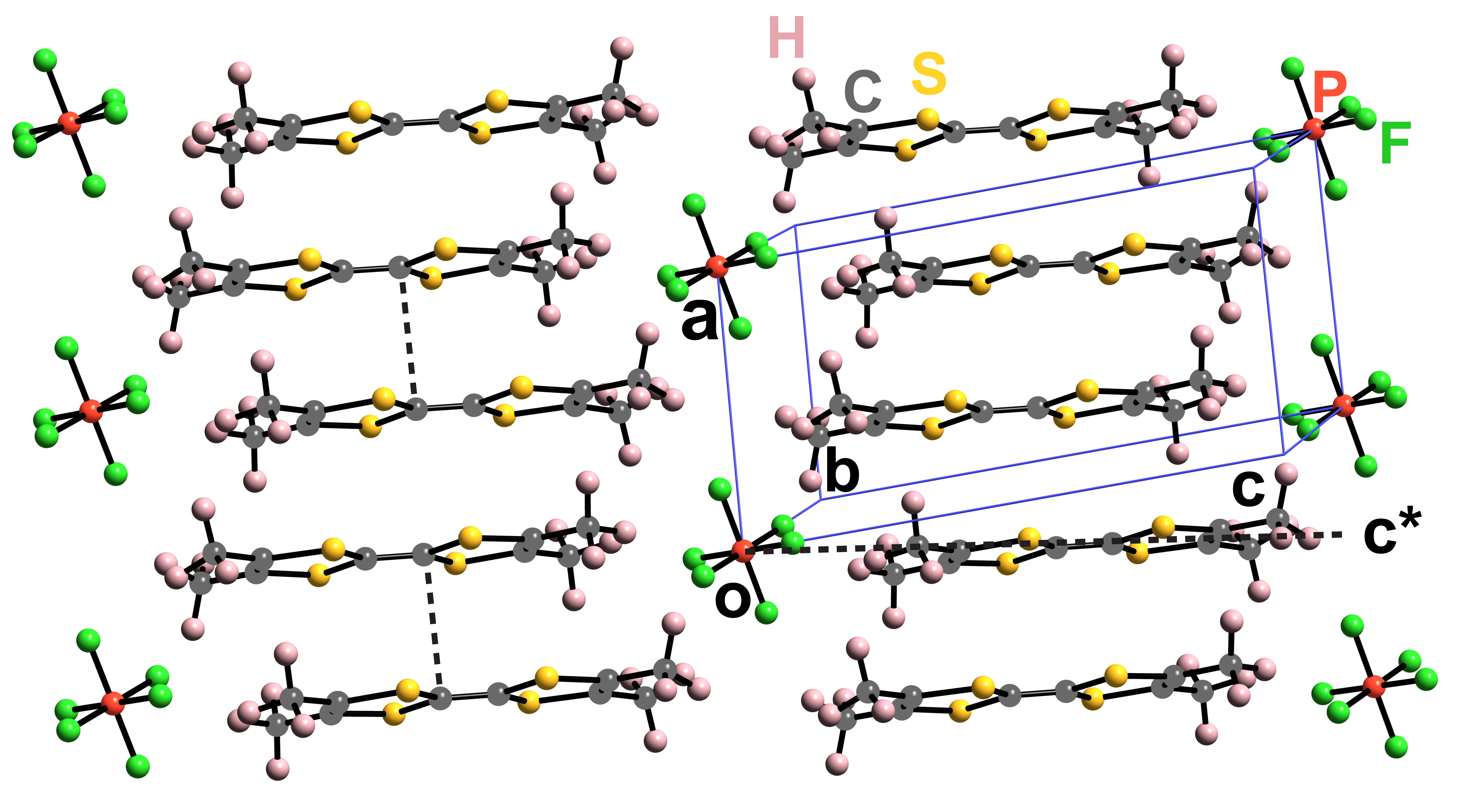}
\caption{\footnotesize Projection of the structure of (TMTTF)$_2X$ with octahedral counter-anion $X$ in the ($a$, $c$) plane. Donors and anions alternate along the $c^*$ direction perpendicular to $a$. Dimers are indicated by the dotted vertical segments in the left stack. Details in the main text.}
\label{Fig-1}
\end{figure}

\begin{figure}
\centering
\includegraphics[angle=0,width=\columnwidth]{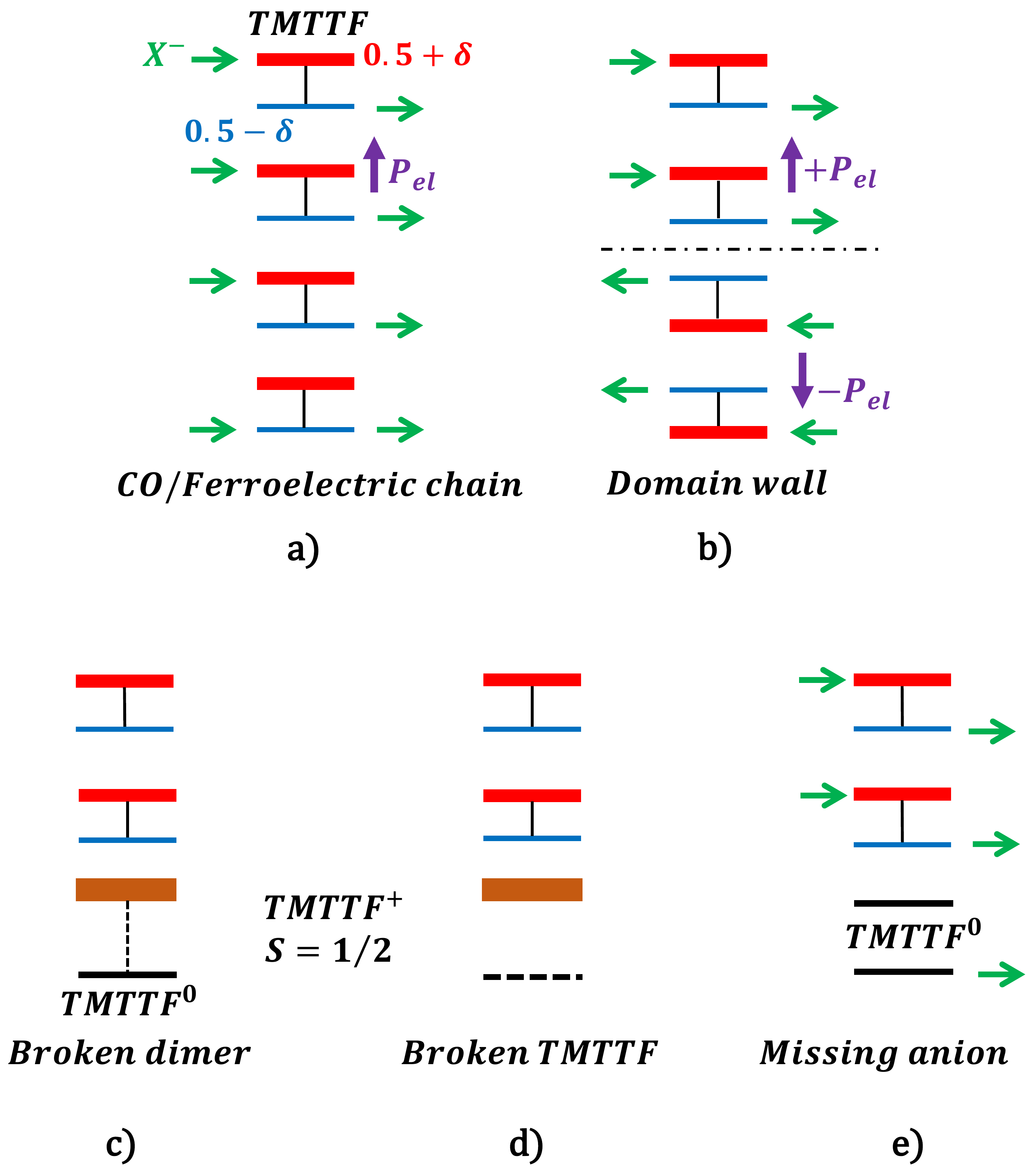}
\caption{\small Scheme of the CO/ferroelectric of the TMTTF chain (label a)) and and its various types of disorder (from label b) to e)). Green arrows represent the anion shift and the vertical black segments the dimer. Charge-rich (poor) donors are in red (blue). TMTTF$^+$ is in brown, TMTTF$^0$ is in black, and the horizontal black dotted segment represents a broken TMTTF.} 
\label{Fig-2}
\end{figure}

Among the organic materials a special attention has been devoted to the Fabre-salts (TMTTF)$_2X$, whose structure is shown in Fig.\,\ref{Fig-1} and where electronic ferroelectricity has been first observed \cite{Monceau} (TMTTF is tetramethyltetrathiafulvalene and $X$ a monovalent counter-anion such as PF$_6$, AsF$_6$ and SbF$_6$). Fabre-salts are composed of well decoupled donor stacks sizeably dimerized in regard of the anions (see Fig.\,\ref{Fig-1}) \cite{newJPref}. In these one-dimensional (1D)  electronic systems, ferroelectricity follows a 1D charge-ordering (CO) transition, revealed by the detection of a charge disproportion ($\pm\,\delta$) between successive donors in the stack direction \cite{ChowPRL2000,DummJ.Phys.IV2004}, occurring at the same temperature as the dielectric constant diverges (below it is noted the common critical temperature $T_{co}$). Both phenomena are driven by electronic correlation effects, which are particularly enhanced in the 1D quarter-filled organic materials built with the TMTTF donor molecule \cite{Bourbon,JPP2012,JPP2015}. As the TMTTF stack is made of TMTTF dimers (outlined in Figs.\,\ref{Fig-1} and \ref{Fig-2}), the charge disproportion breaks all the inversion symmetry of the stack, which bears an electronic polarization ($P_{el}$ in Fig.\,\ref{Fig-2}a)). Since at $T_{co}$ all the stack polarizations are in phase, the material should be a ferroelectric. 
The detection of a strong dielectric response at $T_{co}$, when the electric field is applied along the stack direction $a$, $\epsilon'_a$ $\sim$\,10$^5$ $-$ 10$^6$ in the regime of low frequencies \cite{NadMonceauJPSJ2006,Staresinic} is taken as a good indication that ferroelectricity comes primarily from the electronic sub-system \cite{BrazoII,BrazoIII}. Note that the CO/ferroelectric transition breaks also the symmetry of the magnetic degrees of freedom forming localized $S$ $=$ 1/2 magnetic chains \cite{newJPref2}.

However, since the first detection of a conductivity anomaly at $T_{co}$ \cite{Laversanne1984}, it is found that $T_{co}$ strongly depends upon the nature of the anion $X$ (from $\sim$\,60\,K, 100\,K and 155\,K for $X$ = PF$_6$, AsF$_6$ and SbF$_6$ salts, respectively). This clearly indicates that structural effects has to be incorporated in the CO/ferroelectricity mechanism (for a recent review see, e.g., Ref.\,\cite{PougetCrystal2012}). Also by their easy shift from the inversion centers of the methyl group cavities, anion degrees of freedom are an important ingredient to choose and to stabilize the 3D long-range order formed below $T_{co}$ \cite{BrazoII,BrazoIII}. Such assertion was recently verified by various spectroscopic investigations showing a strengthened interaction of the anion SbF$_6$ \cite{Medjanik2014} and AsF$_6$ \cite{Medjanik2015} with the hole-rich donor molecule TMTTF below $T_{co}$. Nevertheless, evidence for a lattice deformation below $T_{co}$ was only recently obtained by neutron \cite{Pascale2010} and x-ray diffraction \cite{Kitou2017} studies of deuterated and hydrogenated (TMTTF)$_2$PF$_6$ (labelled PF$_6$-D$_{12}$ and PF$_6$-H$_{12}$ hereafter, respectively). In particular, the x-ray study was able to prove the break of inversion symmetry, to determine the amount of charge disproportion between donors and to reveal asymmetric interactions between anions and methyl groups of charge-rich and charge-poor donors (schematically represented by green arrows in Fig.\,\ref{Fig-2}a)).

The pertinence of anion translation and rotation degrees of freedom is assessed by uniaxial lattice thermal expansion measurements exhibiting a sizeable singularity at $T_{co}$, which is more particularly revealed for measurements performed along the interlayer $c^*$-direction \cite{MdS2008,MdS2013} (defined in Fig.\,\ref{Fig-1}). The counter-anions should also control the kinetic of the transition and lead, if $T_{co}$ is low enough, to an incomplete ferroelectric order exhibiting the typical charge response of a relaxor, as observed for the  PF$_6$-H$_{12}$ \cite{Nad2000}. The frequency dependence of the longitudinal dielectric constant $\epsilon'_{a}$ exhibits a smoothing of its divergence together with a shift of its maximum above $T_{co}$ when the frequency of measurement is increased, proving that the pre-transitional dynamics has an intrinsic relaxation character \cite{Review2}. Also, as expected for a typical order-disorder transition, there is a slowing down of the mean-relaxation time at the ferroelectric transition $T_{co}$ \cite{Staresinic}. In this work, we provide evidence that the anion sub-lattice is responsible for such behavior. In this framework the order-disorder type transition can be simply rationalized if it is assumed that each anion is localized in a double-well potential where each minima corresponds to the anion position to form H-bonds with its methyl group environment \cite{PougetCrystal2012}. In presence of such a potential kinetics and dynamics are controlled by the ability of the anion to thermally overcome the double-well potential barrier height in order to change its direction of displacement. Also, anions should be also an intrinsic source of disorder limiting the long-range development of ferroelectricity, which should give rise to a relaxor dielectric response if $T_{co}$ is low enough.

In order to complete previous dielectric measurements performed along the stack direction $a$, we present the first measurements of the dielectric constant ($\epsilon'_{c^*}$) in the transverse inter-layer $c^*$ direction, which is also the direction of contact between anions and the methyl groups of donors. Furthermore, our 1\,kHz low-frequency measurements allows to probe the quasi-static dielectric constant. Thus, $\epsilon'_{c^*}$ measurements allow to investigate directly the charge response of the monovalent counter-anions at the difference of $\epsilon'_{a}$ measurements which should mix electronic and ionic contributions. In addition, we have investigated the effect of controlled disorder on the dielectric charge response by investigating well-characterized \cite{Coulon2007,Coulon2015} x-ray irradiated samples.


\section{Experimental Aspects}
Single crystals of pristine hydrogenated Fabre-salts with the monovalent anions $X$ = PF$_6$, AsF$_6$ and SbF$_6$ were recently synthesized by A. Moradpour$^{\dag}$ in the \emph{Laboratoire de Physique des Solides }(Orsay) employing the standard procedure. Such samples are from the same batches as those investigated in the Electron Spin Resonance (ESR) of Ref.\,\cite{Coulon2007}. The 97.5\% deuterated Fabre-salts with $X$ = PF$_6$ have been also synthesized in Orsay according to the procedure described in Ref.\,\cite{Langlois}. They are of the same batch as those investigated in the microwave dielectric and ESR studies of Refs.\,\cite{Langlois,Coulon2007}, respectively. The samples have a needle-like shape with typical dimensions of (2 $\times$ 1 $\times$ 0.5) mm$^3$ respectively along the $a$, $b'$ (perpendicular to the $a$-axis in the ($a$,$b$) layer) and $c^*$ directions.

Two samples of AsF$_6$-H$_{12}$, three samples of SbF$_6$-H$_{12}$ and PF$_6$-H$_{12}$ and one sample of PF$_6$-D$_{12}$ were measured.  One 12\,h and one 3\,days irradiated AsF$_6$-H$_{12}$ samples corresponding to the ones probed by ESR in Ref.\,\cite{Coulon2007} were also studied.

The x-ray irradiation procedure and the determination of the number of irradiation defects by counting the number of localized spin $S = 1/2$ (see Figs.\,\ref{Fig-2}c) and d)) created as a function of the irradiation time are given in Ref.\,\cite{Coulon2015}. Following the notation used in Ref.\,\cite{Coulon2015}, the number of irradiation defects is expressed in $\%$  of localized spin $S = 1/2$ per mole of TMTTF, a quantity denoted mol$\%$  of irradiation defects. The calibration performed in Ref.\,\cite{Coulon2015} indicates that the 12\,hours irradiated AsF$_6$ sample contains 0.035\,mol$\%$  defect, while the 3\, days irradiated AsF$_6$ sample contains 0.2\,mol$\%$ defect.

In order to perform dielectric measurement along the $c^*$-axis, the ($a$, $b’$) surfaces of each single-crystal were covered with carbon paste supplied by SPI supplies. Tempered gold wires provided by Cryogenic with 0.02\,mm diameter were attached on the surface of the single-crystals already covered with carbon paste.

The capacitance $C$ of the investigated single-crystals was then measured along the $c^*$-axis using the standard two-points probe method employing an Andeen-Hagerling 2550A capacitance bridge with resolution of 10$^{-6}$\,pF, operating in the fixed frequency of 1\,kHz. The quasi-static dielectric constant $\epsilon'$ could be determined by employing the parallel plates capacitor textbook expression, namely $\epsilon =  (C \cdot d)/A$, where $C$ is the capacitance of the sample, $A$ is the capacitor's plates area and $d$ is the distance between them. Hence, the dimensionless dielectric constant $\epsilon' = \epsilon / \epsilon_0$ was obtained, where $\epsilon_0$ is the vacuum permittivity.  A Teslatron-PT cryostat supplied by Oxford Instruments was employed in all experiments in the \emph{Solid State Physics Lab} (Rio Claro).

Before dielectric measurements, the samples were characterized by conductivity measurements along the $c^*$-axis. The $c^*$ resistivity is found to be activated over the whole temperature range. Its thermal dependence, as well as the magnitude of the $c^*$ resistivity, agrees with previous $c^*$ d.c.\,resistivity measurements reported in Ref.\,\cite{Kohler2011}.

Dielectric constant measurements were performed both upon warming and cooling with a temperature rate variation between $\pm$\,6 and $\pm$\,12\,K/h and with an applied electric field between 50 and 150\,mV/cm. Note that our 1\,kHz dielectric measurements provide a good estimate of the quasi-static dielectric constant.

\section{Results and Analysis of the Data}

\subsection{Pristine salts}
\begin{figure}
\centering
\includegraphics[angle=0,width=\columnwidth]{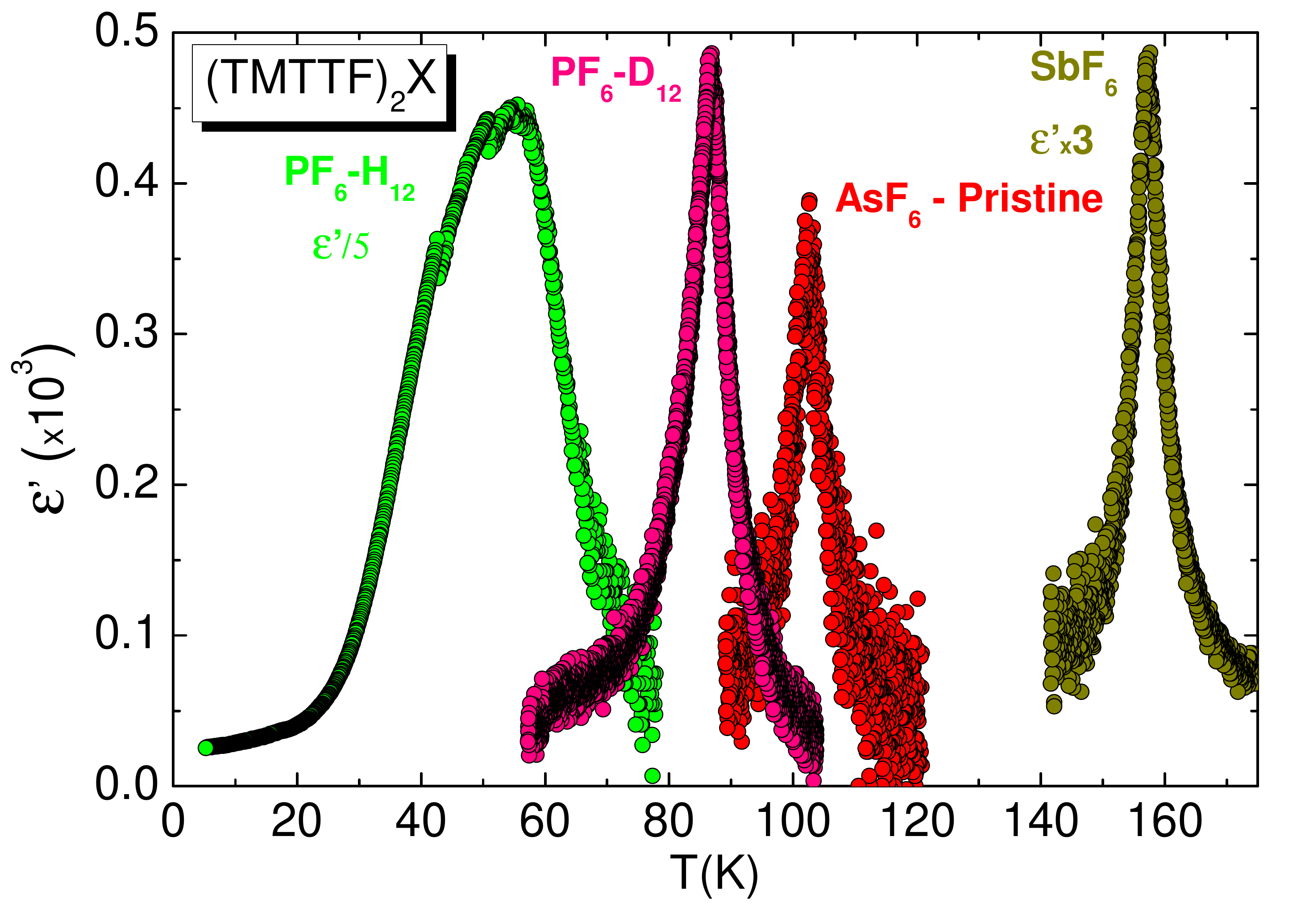}
\caption{\small Temperature dependence of the real part of the $c^*$ dielectric constant $\epsilon'_{c^*}$ of the AsF$_6$-H$_{12}$, SbF$_6$-H$_{12}$, PF$_6$-D$_{12}$ and PF$_6$-H$_{12}$ Sample \#2 salts measured at fixed 1\,kHz upon cooling. For a proper comparison between the various salts, the data for the PF$_6$-H$_{12}$ was divided by a factor of 5 and the SbF$_6$ multiplied by a factor of 3.}
\label{Fig-3}
\end{figure}
Figure\,\ref{Fig-3} shows the thermal dependence of the $c^*$ dielectric constant for the SbF$_6$-H$_{12}$, AsF$_6$-H$_{12}$, PF$_6$-D$_{12}$ and PF$_6$-H$_{12}$ salts measured upon cooling. For all investigated salts, $\epsilon'_{c^*}$  exhibits a well-defined peak-like anomaly at the ferroelectric transition $T_{co}$ of the Fabre-salts, as previously observed for the dielectric constant measured in the stack direction \cite{NadMonceauJPSJ2006}. As can be directly determined from the data set shown in Fig.\,\ref{Fig-3}, the maximum of $\epsilon'_{c^*}$ occurs at $T_{co}$ = 160\,K, 105\,K, 86.6\,K and 57\,K for the SbF$_6$-H$_{12}$, AsF$_6$-H$_{12}$, PF$_6$-D$_{12}$ and PF$_6$-H$_{12}$ salts, respectively. The $T_{co}$ values of SbF$_6$-H$_{12}$ and AsF$_6$-H$_{12}$ are in good agreement with those reported by previous dielectric constant $\epsilon'_{a}$ \cite{NadMonceauJPSJ2006,Staresinic,Nad2000} and local spectroscopic measurements \cite{ChowPRL2000,Yu2004,Dress}.  Also SbF$_6$-H$_{12}$ $T_{co}$ coincides with a net change of slope in the thermal dependence of the $c^*$ resistivity measured on the same sample.
The $\epsilon'_{c^*}$ dielectric measurements on PF$_6$-D$_{12}$ complete an earlier microwave dielectric measurement of $\epsilon'_{a}$ performed in samples of the same batch \cite{Langlois}. The $T_{co}$ value for the PF$_6$-D$_{12}$ salt ($T_{co}$ = 86.6\,K) is close to $T_{co}$ = 84\,K obtained from microwave $\epsilon'_{a}$ and agrees with $T_{co}$ = (84 $\pm$ 3)\,K obtained for the onset of the lattice deformation transition accompanying the CO and detected in an earlier neutron scattering investigation for PF$_6$-D$_{12}$ \cite{Pascale2010}. In PF$_6$-D$_{12}$  of the same batch, an enhancement of the gap of charge is detected at 85\,K by ESR measurements \cite{Coulon2007}.
In the PF$_6$-H$_{12}$, the maximum of $\epsilon'_{c^*}$ observed at 57\,K in Fig.\,\ref{Fig-3} is significantly lower than the $T_{co}$ of 67\,K reported in the literature from local measurements techniques \cite{ChowPRL2000,Dress}.
Note that the thermal dependence of the 1\,kHz dielectric constant $\epsilon'_{a}$ exhibits a maximum at $\sim$\,50\,K \cite{Nad2000}. 

\begin{figure}[htp]

{%
  \includegraphics[clip,width=\columnwidth]{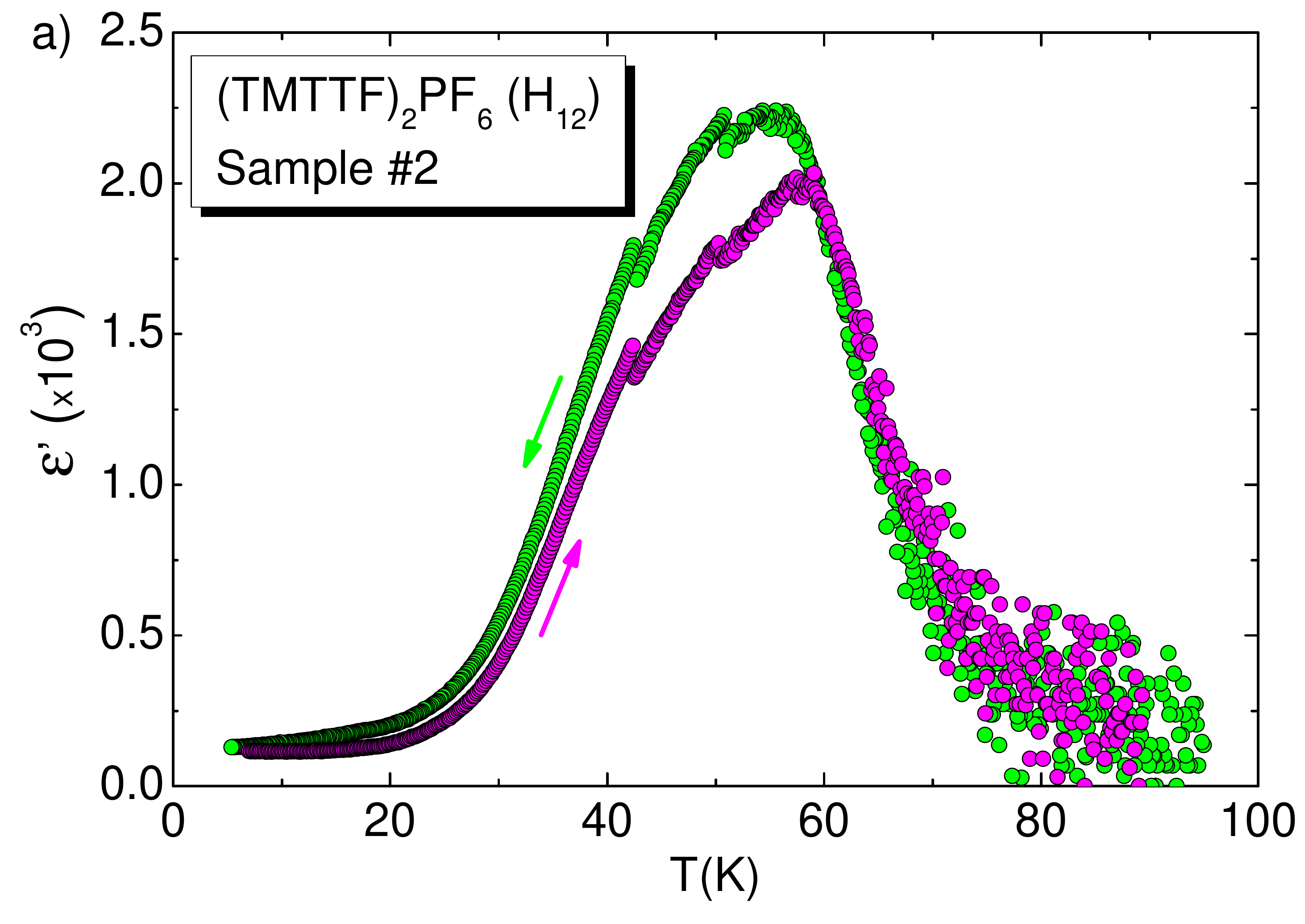}%
}

{%
  \includegraphics[clip,width=\columnwidth]{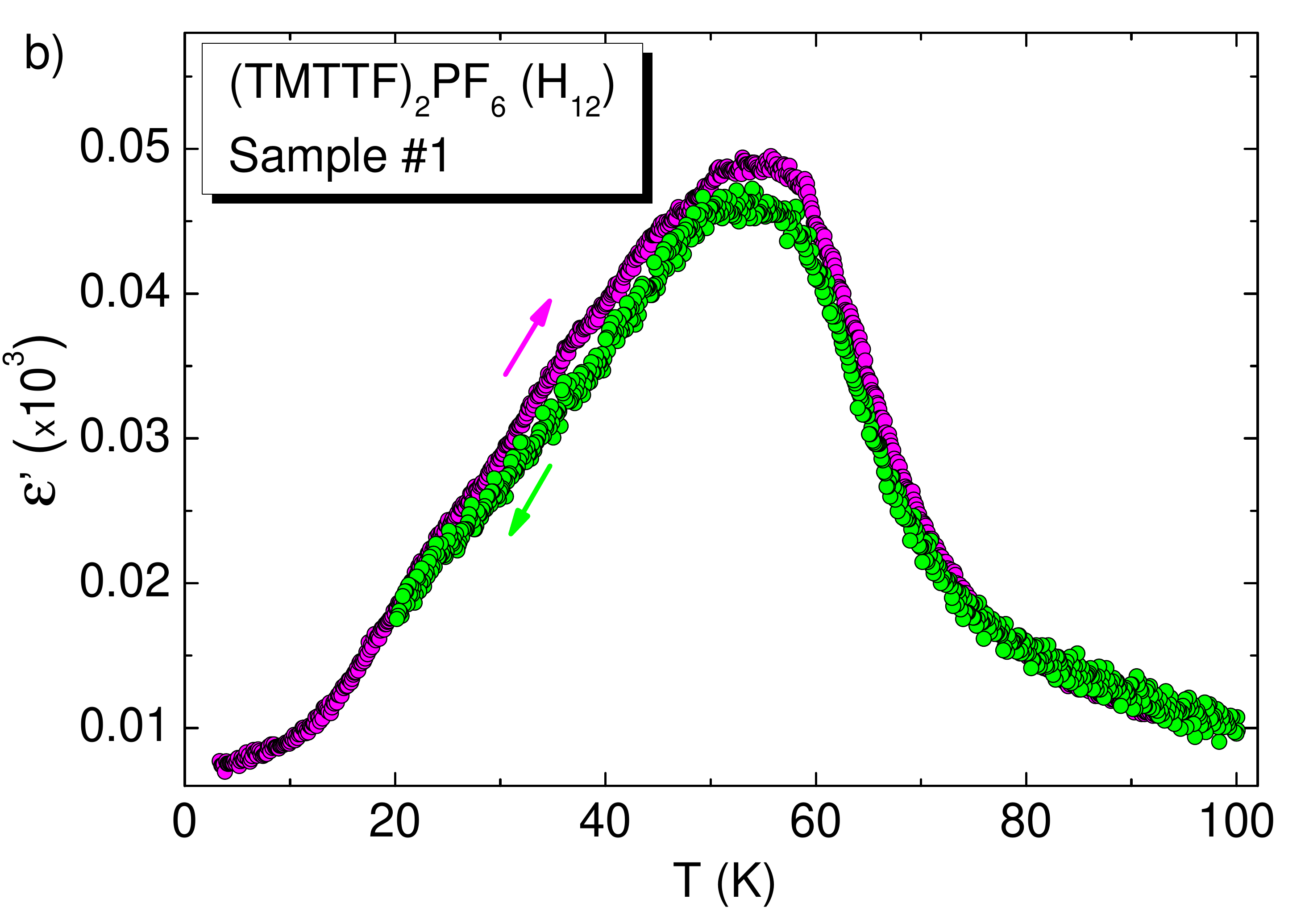}%
}

\caption{a) Dielectric constant as a function of temperature for the fully hydrogenated (TMTTF)$_2$PF$_6$ salt (Sample \#2) showing a maximum dielectric response at about 57\,K both in the heating and cooling processes with employed rate of $\pm$\,7\,K/h and 50\,mV/cm of applied electric  field. Also, the various observed jumps indicate the formation of ferroelectric clusters, details in the main text. b) Dielectric constant as a function of temperature for the fully hydrogenated (TMTTF)$_2$PF$_6$ salt (Sample \#1) showing a broad maximum of dielectric response around 55\,K in both heating and cooling measurements with an employed rate of $\pm$6\,K/h and 150\,mV/cm of applied electric  field.}
\label{Fig-4}
\end{figure}

The amplitude of the maximum of $\epsilon'_{c^*}$  at $T_{co}$ was found to be strongly sample-dependent. The $\epsilon'_{c^*}$  maximum was found to be at:\newline
$-$ 50, 350 and 150 in the three SbF$_6$-H$_{12}$ samples investigated,\newline
$-$ 350 and 150 in the two AsF$_6$-H$_{12}$ samples investigated,\newline
$-$ 400 in the PF$_6$-D$_{12}$ sample investigated.

Unlike the others salts, where a well-defined sharp peak anomaly is observed at $T_{co}$, a broad rounded dielectric response is observed for the PF$_6$-H$_{12}$ salt.  
For Sample \#2 (Fig.\,\ref{Fig-4}a)) a sizeable dielectric response was measured with a main peak-like response up to 2000 at $\sim$\,57\,K followed by two secondary peak-like responses at 40\,K and 52\,K before the vanishing of $\epsilon'_{c^*}$ below $\sim$\,25\,K. For Sample \#1 (Fig.\,\ref{Fig-4}b)) an even broader dielectric response reaching a much weaker intensity of 50 around 55\,K was observed.  Below its broad maximum the response continuously decreases until $\sim$\,10\,K. A behavior similar to the one of Sample \#1 is exhibited by a third sample (not shown here) where $\epsilon'_{c^*}$ exhibits a broad maximum of 90. Note that a somewhat similar broad response strongly frequency-dependent is exhibited by the longitudinal dielectric constant $\epsilon'_{a}$ of PF$_6$-H$_{12}$ investigated in Ref.\,\cite{Nad2000}.

\begin{figure}
\centering
\includegraphics[angle=0,width=\columnwidth]{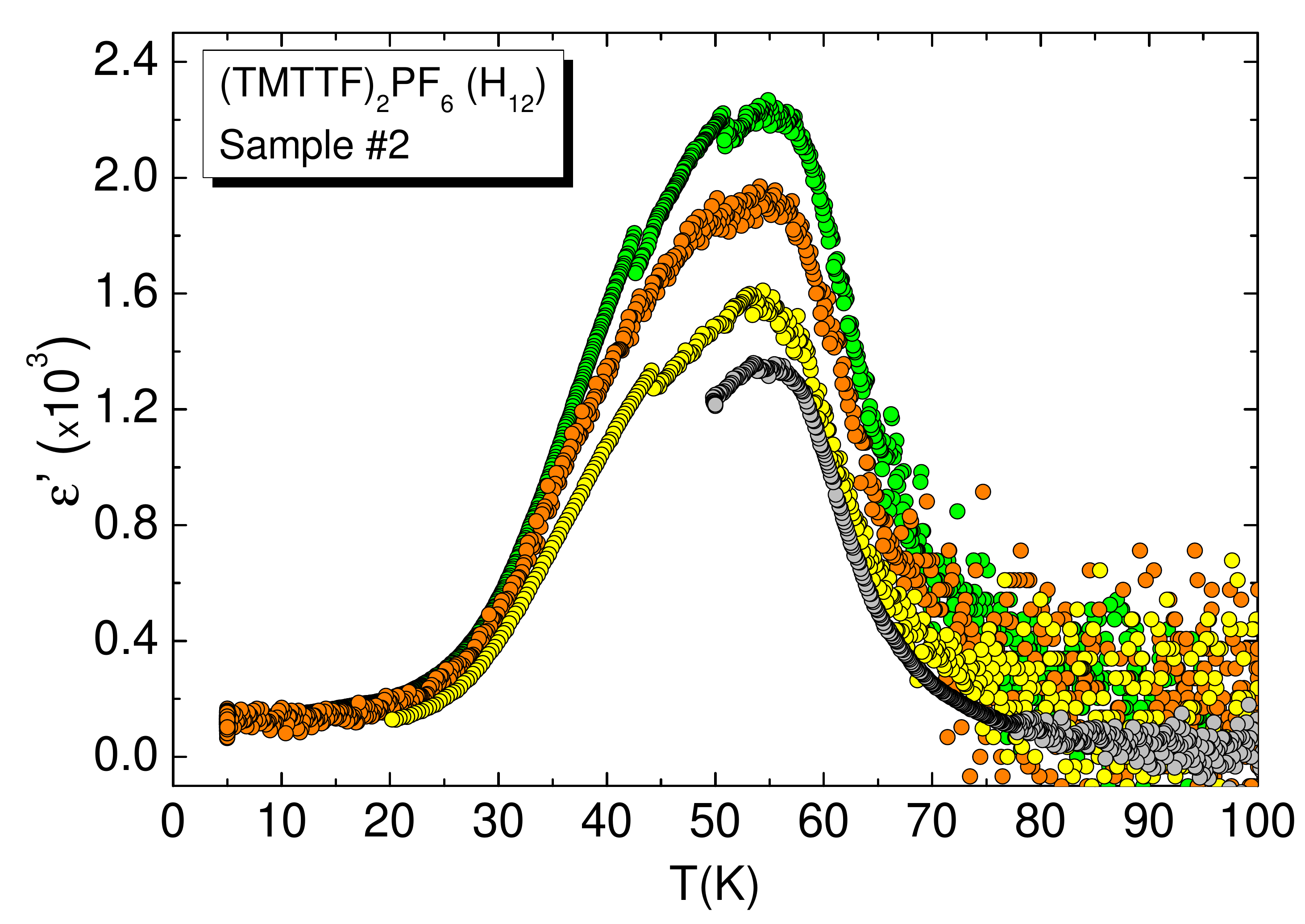}
\caption{\small Dielectric constant of PF$_6$-H$_{12}$ (Sample \#2) measured during four successive cooling runs. The green data set was measured employing a rate of $-$7\,K/h, while for the orange, yellow and gray data sets a rate of $-$10\,K/h was employed.}
\label{Fig-5}
\end{figure}

Figure\,\ref{Fig-4} shows also that there is a significant global hysteresis between $\epsilon'_{c^*}$ of PF$_6$-H$_{12}$ measured upon heating and cooling. Also $\epsilon'_{c^*}$ decreases upon successive thermal cycling (Fig.\,\ref{Fig-5}).

\subsection{Irradiated AsF$_6$-H$_{12}$ salts}
\begin{figure}
\centering
\includegraphics[angle=0,width=\columnwidth]{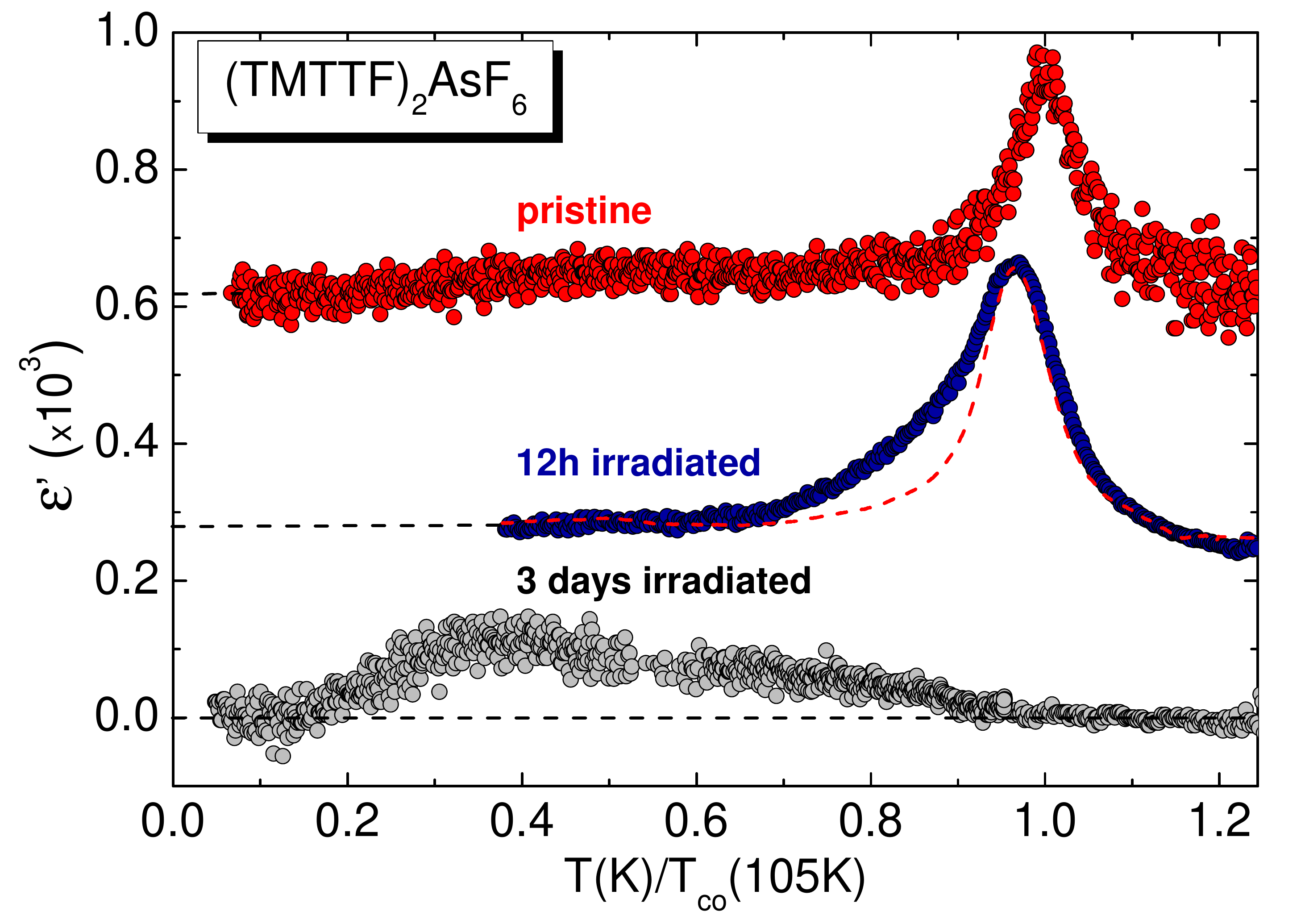}
\caption{\small Dielectric constant ($\epsilon'$) as a function of temperature for the pristine (TMTTF)$_2$AsF$_6$ system (red data set), 12\,h irradiated (blue) and 3 days irradiated (gray). 
The red dashed line, which refers to the smooth data of the pristine variant, was multiplied by a factor of 1.4 for a proper comparison with the 12\,h irradiated variant.  
The measurements were carried out in warming up employing a rate of $+$12\,K/h and with 50\,mV/cm of applied electric field. The temperature scale $T$/$T_{co}$ is expressed in a fraction of the ferroelectric critical temperature of our pristine AsF$_6$ sample ($T_{co}$ = 105\,K). For clarity the data have been shifted vertically (the base line of  each shifted data is indicated on the left side of the figure).}
\label{Fig-6}
\end{figure}

Figure\,\ref{Fig-6} depicts the thermal dependence of $\epsilon'_{c^*}$, normalized to $T_{co}$ of pristine AsF$_6$-H$_{12}$ and for two AsF$_6$-H$_{12}$ samples irradiated for 12\,h and 3\,days. The dielectric response of the irradiated samples broadens pronouncedly  in temperature and shifts to lower temperatures. A comparison between the thermal dependence of pristine and 12\,h irradiated AsF$_6$-H$_{12}$ show that the broadening is basically due to the enhancement of the dielectric response below $T_{co}$. An extreme situation occurs for the 3\,days irradiated AsF$_6$-H$_{12}$ with the growth of a very broad and non-symmetric dielectric response. The latter grows slowly below $T_{co}$, presenting a broad maxima centered at $\sim$\,40\,K, vanishing rapidly below 20\,K.

The $T_{co}$ peak anomaly of pristine AsF$_6$-H$_{12}$ rounds and clearly shifts under irradiation. For the 12\,h irradiated AsF$_6$-H$_{12}$ containing 0.035\,mol\% defect $T_{co}$ was shifted by $\sim$\,3.7\,K. This gives a 1\,K shift of $T_{co}$ for 0.01\,mol\% defect. The sensitivity of the CO transition to irradiation defect is similar to that found in the SbF$_6$ and ReO$_4$ salts, where a 1\,K shift of $T_{co}$ occurs between 0.017mol\,\% defect and 0.007mol\,\% defect respectively\cite{Coulon2015}.

$\epsilon'_{c^*}$ measurements clearly show that irradiation damages change the well-defined peak anomaly of the 1\,kHz (i.e.\,quasi-static) dielectric constant of the pristine salt into a typical broad one of a dielectric relaxor. However, frequency measurements were not performed in order to probe the dynamics of the dielectric response.

\section{Discussion of the results}

\subsection{Anisotropy of the dielectric constant}
All the $c^*$ dielectric measurements presented in Section III show that the transverse dielectric constant $\epsilon'_{c^*}$ is lower by several orders of magnitude than the longitudinal dielectric constant $\epsilon'_{a}$. This proves, as expected for a 1D electronic system, that the charge response is very anisotropic. More quantitatively, if one compares the average transverse $\epsilon'_{c^*}$ measured at 1\,kHz with the longitudinal $\epsilon'_{a}$ measured at 10\,kHz in  Ref.\,\cite{Staresinic} for the AsF$_6$-H$_{12}$, one obtains a $\epsilon'_{a}$/$\epsilon'_{c^*}$ ratio of 400 at $T_{co}$. A similar $\epsilon'_{a}$/$\epsilon'_{c^*}$ ratio of $\sim$\,10$^3$ at $T_{co}$ is obtained in PF$_6$-H$_{12}$ from the ratio of the 1\,kHz longitudinal dielectric constant of Ref.\,\cite{Nad2000} to the average transverse dielectric constant. In these salts the anisotropy of $\epsilon'$ is comparable to the anisotropy of electrical conductivity between $a$ and $c^*$ directions at $T_{co}$ \cite{Kohler2011}.

Among the ionic species a special attention must be devoted to the anions, which are the most mobile charged constituents of the Fabre-salts and thus should exhibit a strong dielectric response \cite{Review2}. $^{19}$F NMR probe of SbF$_6$-H$_{12}$ and SbF$_6$-D$_{12}$ salts shows that the anion begins to rotate in their methyl group cavity above $\sim$\,75$-$80\,K \cite{Yu2004} and undergoes successive reorientation with thermally activated jumps movements above 135 and 210\,K \cite{Furukawa2005}. In addition, rotation and translational degrees of freedom of the anions are taken as responsible for the negative thermal expansion of lattice parameters detected above $\sim$\,100\,K, this onset temperature being slightly dependent on the nature of the anion, see Refs.\,\cite{MdS2013,MdS2008}. Such anomalous behavior is found to be the strongest in the $c^*$ direction, along which the donors and anions layers alternate. Hence, dielectric measurements performed in the $c^*$ direction should mostly probe the anion contribution at the dielectric response. The detection of a critical divergence of $\epsilon'_{c^*}$ at $T_{co}$, which will be more quantitatively analyzed in the next section, means that anions collectively participate at the onset of the ferroelectric ground-state. In particular, its displacement from the inversion centers of the high temperature lattice is essential to break the inversion lattice symmetry and to stabilize the 3D electronic charge-pattern below $T_{co}$.

Note that it is also expected, due to the isotropic nature of the anion cavity, an ionic contribution at $\epsilon'_{a}$ which should be of the same order of magnitude as $\epsilon'_{c^*}$. Nevertheless, if the ionic contribution scales with the anion mobility giving rise to the anomalous thermal dependence of the lattice expansion coefficient above $\sim$\,100\,K\cite{MdS2013,MdS2008}, one expects a larger ionic contribution for $\epsilon'_{c^*}$.

\subsection{Thermal dependence of the dielectric response}
Figure\,\ref{Fig-3} clearly shows that the thermal broadening of the quasi-static dielectric response $\epsilon'_{c^*}$ of the different Fabre-salts increases upon decreasing $T_{co}$. More quantitatively, if one defines $\Delta T$ as the full-width at half-maximum temperature range of the dielectric response, one gets $\Delta T$/$T_{co}$ of 3\%, 5.5\% and 8\% for the SbF$_6$-H$_{12}$, AsF$_6$-H$_{12}$ and PF$_6$-D$_{12}$ salts, respectively. Much larger $\Delta T$/$T_{co}$ values of 0.47 and of 0.66 are obtained for PF$_6$-H$_{12}$ Sample \#2 and \#1, cf. Figs.\,\ref{Fig-4}a) and 4b), respectively.

Except for the PF$_6$-H$_{12}$, this systematic broadening effect is less apparent in $\epsilon'_{a}$ measured at higher frequency for the various Fabre-salts investigated in Ref.\,\cite{NadMonceauJPSJ2006}. This observation indicates that $\epsilon'_{c^*}$, which mostly probes the anion response, is more sensitive to the structural disorder. Such observation is not surprising as the anions themselves are the source of the disorder. In order to test this assertion, the dielectric response of irradiated AsF$_6$-H$_{12}$ was similarly analyzed. Figure\,\ref{Fig-6} shows that their dielectric response significantly broadens upon increasing irradiation. More quantitatively, one obtains $\Delta T$/$T_{co}$ of 0.15 and 1.25 for AsF$_6$-H$_{12}$ irradiated 12\,hours (0.035\,mol\% defect) and 3\,days (0.2\,mol\% defect), respectively.

If one assumes that the $\Delta T$/$T_{co}$ broadening is roughly proportional to the amount of defect, one estimates that PF$_6$-H$_{12}$ intrinsically contains the equivalent of $\sim$\,0.1\,mol\% defect. This is 5 times larger than the estimated amount of intrinsic defects in PF$_6$-D$_{12}$, which should contain the equivalent of $\sim$\,0.02\,mol\% defect.

\begin{figure}[htp]

{%
  \includegraphics[clip,width=\columnwidth]{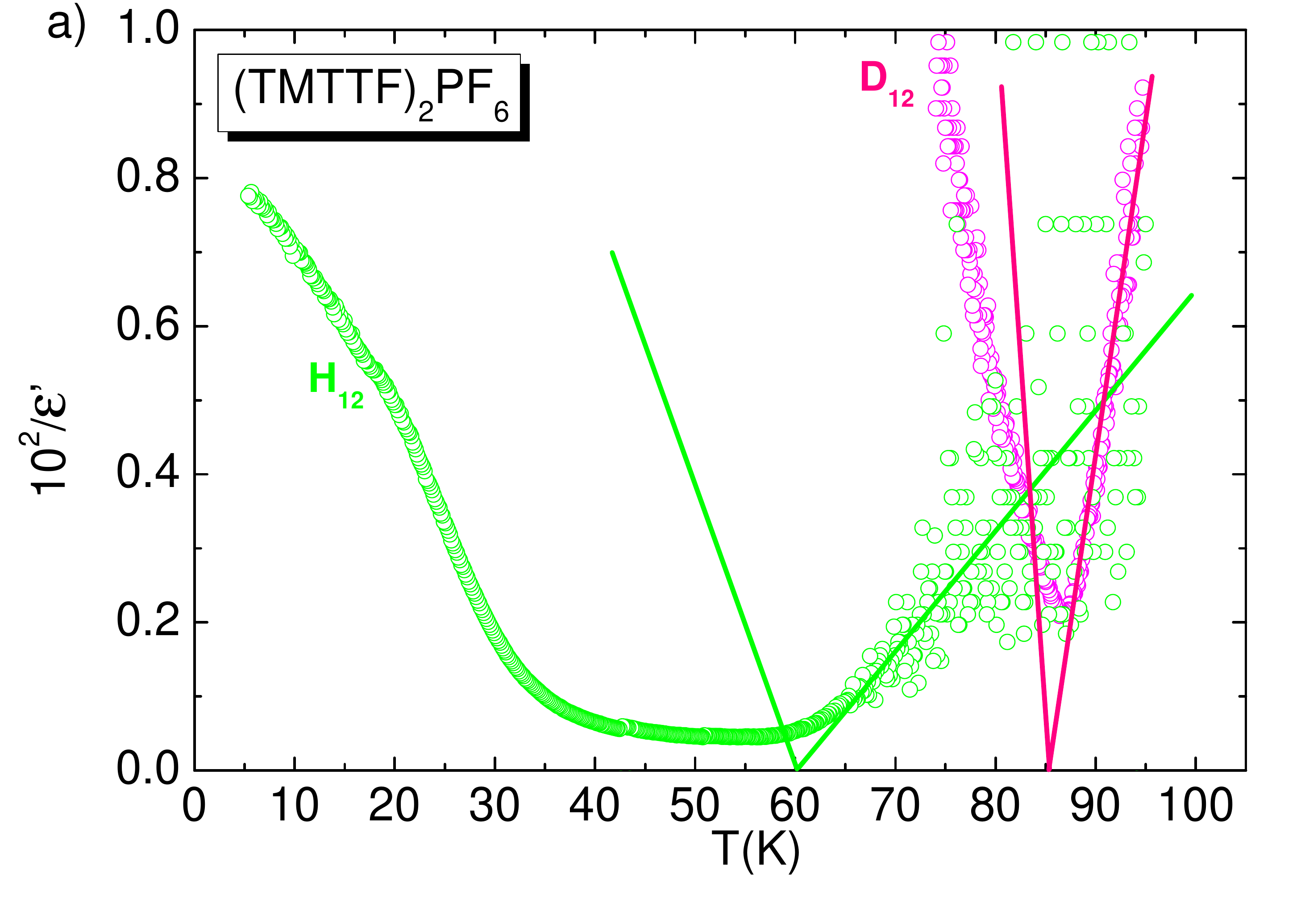}%
}

{%
  \includegraphics[clip,width=\columnwidth]{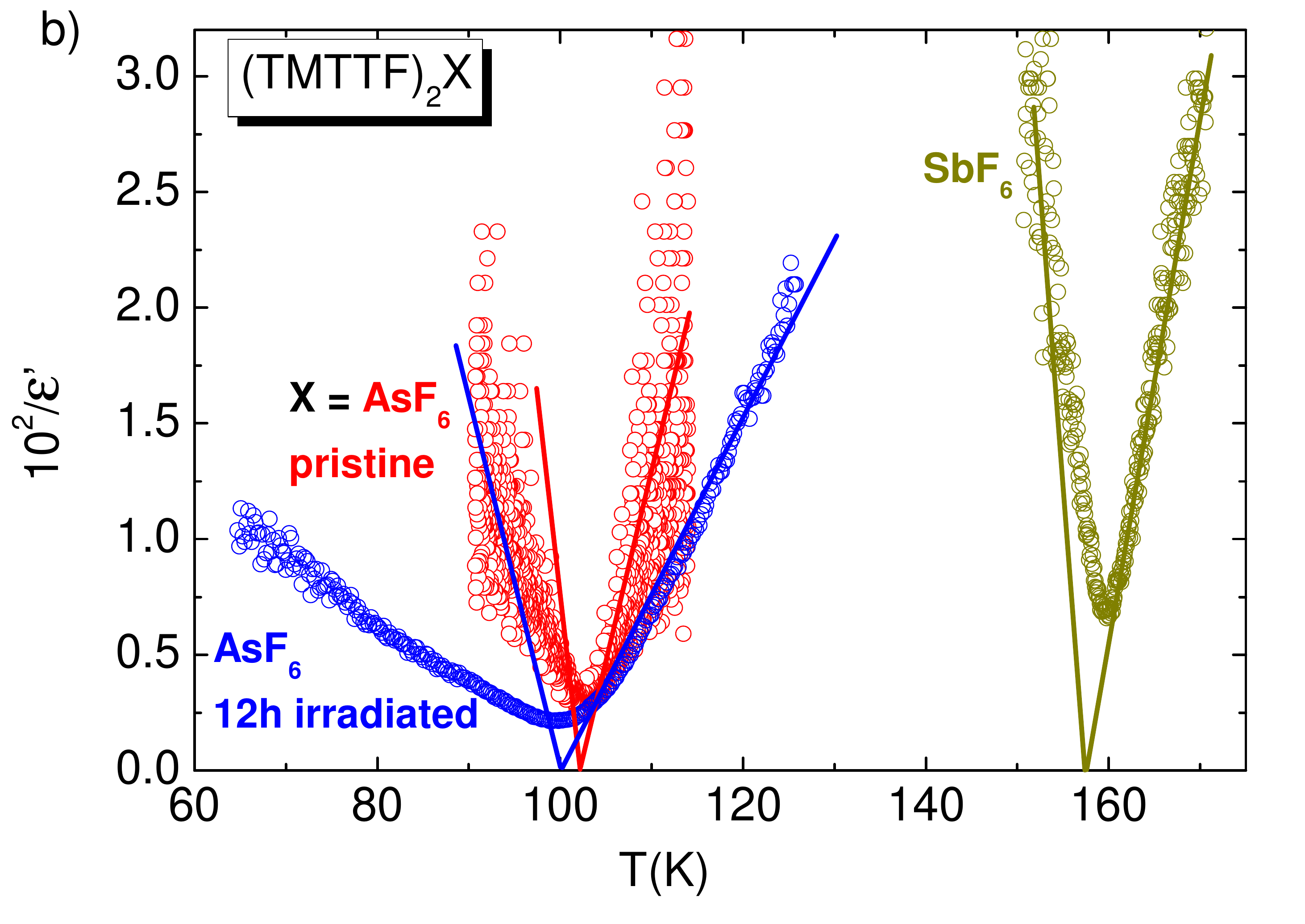}%
}

\caption{Curie-Weiss plot, i.e., 1/$\epsilon'$ as a function of $T$ upon cooling for: a) H$_{12}$ Sample \#2 (green) and D$_{12}$  (pink) PF$_6$; b) the pristine (TMTTF)$_2$AsF$_6$ system (red data set), 12\,hours irradiated (blue), SbF$_6$ (dark yellow), cf.\,label. The solid lines indicate the expected slopes in the frame of the mean-field approximation. The solid lines above $T_{co}$ are obtained from a mean-square fit of the raw data for the various salts. Details in the main text.}
\label{Fig-7}
\end{figure}

The thermal divergence of $\epsilon'_{c^*}$ at $T_{co}$ can be quantitatively analyzed by the Curie-Weiss plots of 1/$\epsilon'_{c^*}$ as a function of $T$ (Figs.\,\ref{Fig-7}a) and b)). In the mean-field approximation it is predicted that in both sides of $T_{co}$:

\begin{equation*}
1/\epsilon'_{c^*} = A_{\pm} | T - T_{co} |,
\label{factor2}
\end{equation*}

\noindent with a slope ratio $A_{-}$/$A_{+}$ of 2. The criteria employed for the Curie-Weiss analysis, depicted in Figs.\,\ref{Fig-7}a) and \ref{Fig-7}b), was to fix the right solid line, obtained from a mean least square fit of the raw data above $T_{co}$, and draw the left solid line in accordance with the slope ratio in the mean-field approximation ($A_{-}$/$A_{+}$ = 2), as shown in Ref.\,\cite{Staresinic}. This mean-field dependence is only observed for 1/$\epsilon'_{c^*}$ of the SbF$_6$-H$_{12}$. In the other Fabre-salts one observes a deviation from the linear dependence below $T_{co}$, which increases upon going from AsF$_6$-H$_{12}$, PF$_6$-D$_{12}$ to PF$_6$-H$_{12}$. In contrast, this linear behavior is found in both sides of $T_{co}$ for 1/$\epsilon'_{a}$ reported in Refs.\,\cite{Monceau,Nad2000} in SbF$_6$-H$_{12}$, AsF$_6$-H$_{12}$ and PF$_6$-H$_{12}$. However, a deviation of the linear dependence of 1/$\epsilon'_{a}$ below $T_{co}$, similar to the one exhibited by 1/$\epsilon'_{c^*}$ in Fig.\,\ref{Fig-7}, is found in a recent reinvestigation of $\epsilon'_{a}$ of AsF$_6$-H$_{12}$ \cite{Staresinic}.

Figure\,\ref{Fig-7} shows that the deviation from the linear dependence of 1/$\epsilon'_{c^*}$ below $T_{co}$ is dramatically enhanced in the 12\,hours irradiated AsF$_6$-H$_{12}$ sample. Such a behavior is also visible in the $\epsilon'_{c^*}$ measurements shown in Fig.\,\ref{Fig-6}. Based on such observation, one can thus ascribes the deviation from the mean-field divergence of the dielectric constant to the formation of structural disorders, regardless of their origin. A similar conclusion was achieved from the study of $\epsilon'_{a}$ in AsF$_6$-H$_{12}$ \cite{Staresinic}.

\subsection{Sources of disorder}
Figures\,\ref{Fig-2}b) to \ref{Fig-2}e) give schematic illustrations of the various sources of disorder, which can be present in the Fabre-salts. As a matter of fact, recent Raman scattering investigations have detected ionized TMTTF$^+$ and neutral TMTTF$^0$ species in pristine SbF$_6$-H$_{12}$, AsF$_6$-H$_{12}$ and PF$_6$-H$_{12}$ \cite{Swietlik2017}, which were assigned to ferroelectric domain-walls. Since in a dimer the average charged donor is TMTTF$^{\textmd{+0.5}}$, neutral and ionized species indicates the existence of broken dimers (Fig.\,\ref{Fig-2}c)). It is known that the photo-induced local chemistry caused by x-ray irradiation leads to the break of both donor and anion species \cite{Coulon2015}. This also induces molecular displacements in the surroundings of a defect. The break of a donor molecule should prevent the pairing of donors into dimers, leading to the formation of TMTTF$^+$ species (Fig.\,\ref{Fig-2}d)). The break of an anion in turn will prevent the transfer of one electron from the TMTTF stack to the anion. This will provide TMTTF$^0$ species instead of two TMTTF$^{\textmd{+0.5}}$, which usually pair into a dimer (in Fig.\,\ref{Fig-2}e) there are two TMTTF$^0$).

Another important source of disorder arise from the reversal direction of displacement of the anions. As shown in Fig.\,\ref{Fig-2}b), this will interchange the position of charge-rich and charge-poor donors along the stack direction. Such interchange can be viewed as the introduction of a phase shift of $\pi$ in the CO stack modulation \cite{Brazo}. This phase shift changes the direction of the electronic polarization, introducing a ferroelectric domain-wall. As the transverse displacement of the anions fixes the longitudinal direction of the electronic polarization, and vice-versa, such defects should affect both $\epsilon'_{c^*}$ and $\epsilon'_{a}$.

Except for the irradiated samples, where TMTTF$^+$ and TMTTF$^0$ defects (Figs.\,\ref{Fig-2}d) and \ref{Fig-2}e)) are directly created by the x-ray irradiation local chemistry, it appears that a change of direction in the uniform anion shift should be easily achieved in the pristine Fabre-salts. The presence of such defects is most likely frequent, since the anions are less mobile and begin to freeze in the methyl group cavities when the temperature is decreased. For the SbF$_6$-H$_{12}$ ($T_{co}$ = 160\,K), the ferroelectric transition occurs in the temperature range, where the presence of mobile anions is assessed by NMR \cite{Yu2004,Furukawa2005} together with a $c^*$ negative lattice thermal expansion above $\sim$\,80\,K \cite{MdS2013}.  Thus, SbF$_6$-H$_{12}$ is not expected to exhibit important disorder effects, modifying the Curie-Weiss dependence of the dielectric constant. This is not the case for the AsF$_6$-H$_{12}$ ($T_{co}$ = 105\,K) because the ferroelectric transition occurs when the anions just begin to freeze with the saturation of the $c^*$ negative lattice thermal expansion regime around 125\,K \cite{MdS2013,JPP2015}. Compared to the SbF$_6$-H$_{12}$ salt, a longer tail of the dielectric response develops below $T_{co}$ in the AsF$_6$-H$_{12}$ salt. A somewhat similar observation can be done for the PF$_6$-D$_{12}$ salt ($T_{co}$ = 86.6\,K). A very drastic effect occurs in the PF$_6$-H$_{12}$ salt, since $T_{co}$ occurs below the temperature $\sim$\,75\,K at which the anion movements freeze, according to the $c^*$ thermal expansion measurements reported in Refs.\,\cite{MdS2013,MdS2008}.

The thermally broadened dielectric response of PF$_6$-H$_{12}$ exhibits clear characteristics of a ferroelectric relaxor \cite{Samara2003}. First, the dielectric response is broad and does not exhibit a clear peak-like anomaly at a well-defined $T_{co}$. Second, the maxima of $\epsilon'$ depends on the frequency of the measurement \cite{Nad2000}. In the low-frequency regime, i.e. in quasi-static conditions, the maxima of $\epsilon'_{a}$ occurs at $\sim$\,50\,K \cite{Nad2000} and the maxima of $\epsilon'_{c^*}$ occurs around 55\,K and 57\,K for the two samples measured in Fig.\,\ref{Fig-4}. All the $T_{co}$ values obtained from the maximum of $\epsilon'$ are significantly lower than $T_{co}$ = 67\,K obtained from local detection, i.e.\,from spectroscopic techniques \cite{ChowPRL2000,Dress}. Our $\epsilon'_{c^*}$ measurements show also that the charge response is sample dependent. A broad response with a low value of $\epsilon'_{c^*}$ is observed in Sample \#1 (Fig.\,\ref{Fig-4}b)), while a narrower response of a larger $\epsilon'_{c^*}$ accompanied by several secondary peaks is observed in Sample \#2 (Fig.\,\ref{Fig-4}a)). This behavior suggests that Sample \#2 is composed of ferroelectric clusters with different $T_{co}$. Third, one observe a global hysteresis between the heating and cooling curves (Fig.\,\ref{Fig-4}), and a reduction of $\epsilon'_{c^*}$ upon successive cycling (Fig.\,\ref{Fig-5}). The global hysteresis point out the presence of disorder, which pins ferroelectric domains and prevents their thermal evolution. Each cycle creates defects, which  limit the extent of ferroelectric domains in the subsequent cooling processes because they are not completely resorbed after the heating cycle above $T_{co}$.

In agreement with the observation of a slowing down of the mean-relaxation time at $T_{co}$ \cite{Staresinic}, it takes a long time to establish the long-range thermodynamic ferroelectric order. Thus, the slow kinetics associated with the anion displacement process  limit the ferroelectric ordering process. Then, the difficulties to achieve the long-range uniform anion shift should be taken as responsible for the fragmentation of the ferroelectric order into domains. If one assumes that PF$_6$-H$_{12}$ contains 0.1\,mol$\%$ defect, there is one defect every 1000 TMTTF molecules or every 500 unit cells (there are two TMTTF per unit cell). If one defect pins one ferroelectric domain, the average volume of the domain should be of (3.5 $\times$ 10$^5$)\,$\textmd{\AA}^3$. 
This volume is close to the average volume of the CO domain, namely (8.7 $\times$ 10$^5$)\,$\textmd{\AA}^3$, estimated for irradiated (TMTTF)$_2X$ \cite{Coulon2015} from the defect concentration at which the singular decay of conductivity at the CO transition vanishes. If the ferroelectric domains pinned on defects are isotropic, its average size is about 70\,$\textmd{\AA}$.


\vspace{0.5cm}

\section{Conclusion}
We have measured the quasi-static transverse dielectric constant of the Fabre-salts allowing to probe the ionic charge response, which is the lattice counterpart of the electronic ferroelectricity associated with the CO transition. Measurements along the $c^*$ direction give directly access to the counter-anion response, which is a key ingredient to choose and to stabilize the 3D long-range order below $T_{co}$. Our $\epsilon'_{c^*}$ measurements, conjugated with earlier investigations of the lattice thermal expansion, show that the dielectric response is thermally broaden below $T_{co}$ if the transition occurs in the temperature range where the anion movement begin to freeze in the methyl group cavities. In the case of PF$_6$-H$_{12}$, where $T_{co}$ occurs in the freezing regime, a relaxor-type ferroelectricity is observed. For the latter, because of the slow kinetics of the anion sub-lattice, global hysteresis effects and reduction of the charge response upon successive thermal cycling is observed. Finally, $\epsilon'_{c^*}$ measurements of x-ray irradiated AsF$_6$-H$_{12}$ show that irradiations damages change the well-defined dielectric response of the pristine salts into a relaxor one. In this context, we have proposed that anions control the order-disorder or relaxation character of the ferroelectric transition of the Fabre-salts.

\section{Acknowledgements}
M.\, de S.\, acknowledges financial support from the S\~ao Paulo Research Foundation -- Fapesp (Grants No. 2011/22050-4), National Council of Technological and Scientific Development -- CNPq (Grants No.\,305472/2014-3) and CAPES for granting the M. Sc. scholarships for L. S. and C. S..

\medskip

\bibliographystyle{unsrt}
\bibliography{References}

\begin{thebibliography}{10}

\bibitem{Lebed}
Physics of Organic Conductors and Superconductors, edited by A.G. Lebed,
  Springer Series in Materials Sciences (Springer-Verlag, Berlin, Heidelberg,
  2008), Vol. 110.

\bibitem{reviewdressel}
M. Dressel, \emph{Naturwissenschaften} \textbf{94}, 527 (2007).

\bibitem{MdS2013}
M. de Souza and J.-P. Pouget, J. Phys.: Condens. Matter \textbf{25}, 343201
  (2013).

\bibitem{Review1}
S. Tomic and M. Dressel, Rep. Prog. Phys. \textbf{78}, 096501 (2015).

\bibitem{Review2}
P. Lunkenheimer and A. Loidl, J. Phys.: Condens. Matter \textbf{27}, 373001
  (2015).

\bibitem{Ishihara}
Sumio Ishihara, J. Phys.: Condens. Matter 26, 493201 (2014).

\bibitem{Brazo}
S. Brazovskii in Ref. [1] p. 313.

\bibitem{BrazoII}
S. Brazovskii, J. Phys. IV France \textbf{12}, Pr9-149 (2002).

\bibitem{BrazoIII}
S. Brazovskii, Synth. Metals \textbf{133-134}, 301 (2003).

\bibitem{Monceau}
P. Monceau, F. Ya Nad, S. Brazovskii, Phys. Rev. Lett. \textbf{86}, 4080
  (2001).

\bibitem{newJPref}
J.-P. Pouget and S. Ravy, J. Phys. I France \textbf{6}, 1501 (1996).

\bibitem{ChowPRL2000}
D. S. Chow, F. Zamborszky, B. Alavi, D. J. Tantillo, A. Baur, C. A. Merlic, and
  S. E. Brown, Phys. Rev. Lett. \textbf{85}, 1698 (2000).

\bibitem{DummJ.Phys.IV2004}
M. Dumm, B. Salameh, M. Abaker, L. K. Montgomery, and M. Dressel,J. Phys. IV
  (France) \textbf{114}, 57 (2004).

\bibitem{Bourbon}
C. Bourbonnais and D. J\'erome in Ref. [1] p. 357.

\bibitem{JPP2012}
J.-P. Pouget, Physica B \textbf{407}, 1762 (2012).

\bibitem{JPP2015}
J.-P. Pouget, Physica B \textbf{460}, 45 (2015).

\bibitem{NadMonceauJPSJ2006}
F. Nad and P. Monceau, J. Phys. Soc. Japan \textbf{75}, 051005 (2006).

\bibitem{Staresinic}
D. Stare\v{s}ini\'c, K. Biljakovi\'c, P. Lunkenheimer and A. Loidl, Solid State
  Commun. \textbf{137}, 241 (2006).

\bibitem{newJPref2}
S. Yasin, B. Salameh, E. Rose, M. Dumm, H.-A. Krug von Nidda, A. Loidl, M.
  Ozerov, G. Untereiner, L. Montgomery, and M. Dressel, Physical Review B
  \textbf{85}, 144428 (2012).

\bibitem{Laversanne1984}
R. Laversanne, C. Coulon, B. Gallois, J. P. Pouget, and R. Moret, J. Phys.
  (Paris) Lett. \textbf{45}, L393 (1984).

\bibitem{PougetCrystal2012}
J.-P. Pouget, Crystals \textbf{2}, 466 (2012).

\bibitem{Medjanik2014}
K. Medjanik, M. de Souza, D. Kutnyakhov, A. Gloskovskii, J. Muller, M. Lang,
  J.-P. Pouget, P. Foury-Leylekian, A. Moradpour, H.-J. Elmers, \emph{et al.}
  Eur. Phys. J. B \textbf{87}, 256 (2014).

\bibitem{Medjanik2015}
K. Medjanik, A. Chernenkaya, S. A. Nepijko, G. Ohrwall, P. Foury-Leylekian, P.
  Alemany, E. Canadell, G. Schonhense, J.-P. Pouget, Phys. Chem. Chem. Phys.
  \textbf{17}, 19202 (2015).

\bibitem{Pascale2010}
P. Foury-Leylekian, S. Petit, G. Andre, A. Moradpour and J.-P. Pouget, Physica
  B \textbf{405}, S95 (2010).

\bibitem{Kitou2017}
S. Kitou, T. Fujii, T. Kawamoto, N. katayama, S. Maki, E. Nishibori, K.
  Sugimoto, M. Takata, T. Nakamura and H. Sawa, Phys. Rev. Lett. \textbf{119},
  065701 (2017).

\bibitem{MdS2008}
M. de Souza, P. Foury-Leylekian, A. Moradpour, J.-P. Pouget and M. Lang, Phys.
  Rev. Lett. \textbf{101}, 216403 (2008).

\bibitem{Nad2000}
F. Nad, P. Monceau, C. Carcel, J. M. Fabre, Phys. Rev. B \textbf{62}, 1753
  (2000).

\bibitem{Coulon2007}
C. Coulon, G. Lalet, J.-P. Pouget, P. Foury-Leylekian, A. Moradpour and J. M.
  Fabre, Phys. Rev. B \textbf{76}, 085126 (2007).

\bibitem{Coulon2015}
C. Coulon, P. Foury-Leylekian, J. M. Fabre and J.-P. Pouget, Eur. Phys. J. B
  \textbf{88}, 85 (2015).

\bibitem{Langlois}
A. Langlois, M. Poirier, C. Bourbonnais, P. Foury-Leylekian, A. Moradpour and
  J.-P. Pouget, Phys. Rev. B \textbf{81}, 125101 (2010).

\bibitem{Kohler2011}
B. K$\ddot{\textmd{u}}$hler, E. Rose, M. Dumm, G. Untereiner, and M. Dressel,
  Phys. Rev. B \textbf{84}, 035124 (2011).

\bibitem{Yu2004}
W. Yu, F. Zhang, F. Zamborszky, B. Alavi, A. Baur, C. A. Merlic, and S. E.
  Brown, Phys. Rev. B \textbf{70}, 121101(R) (2004).

\bibitem{Dress}
M. Dressel, M. Dumm, T. Knoblauch, M. Masino, Crystals \textbf{2}, 528 (2012).

\bibitem{Furukawa2005}
K. Furukawa, T. Hara, and T. Nakamura, J. Phys. Soc. Jpn. 7\textbf{4}, 3288
  (2005).

\bibitem{Swietlik2017}
R. \ifmmode \acute{S}\else \'{S}\fi{}wietlik, B. Barszcz, A. Pustogow, M.
  Dressel, Phys. Rev. B \textbf{95}, 085205 (2017).

\bibitem{Samara2003}
G. A. Samara, J. Phys.: Condens. Matter \textbf{15}, R367 (2003).

\end{thebibliography}
\end{document}